\documentclass[aps,prx,reprint,superscriptaddress,amsmath,amssymb,]{revtex4-2}

\usepackage{graphicx}
\usepackage{dcolumn}
\usepackage{bm}

\usepackage{braket}
\usepackage{float}
\usepackage[colorlinks=true,urlcolor=blue,linkcolor=blue,citecolor=blue]{hyperref}
\usepackage{physics}
\usepackage{siunitx}

\begin{document}

\preprint{Enhancing Dispersive Readout of Superconducting Qubits Through Dynamic Control of the Dispersive Shift: Experiment and Theory}

\title{Enhancing Dispersive Readout of Superconducting Qubits Through Dynamic Control of the Dispersive Shift: Experiment and Theory
}

\author{Fran\c cois~Swiadek}
    \email{francois.swiadek@phys.ethz.ch}
    \affiliation{Department of Physics, ETH Zurich, 8093 Zurich, Switzerland}
    \affiliation{Quantum Center, ETH Zurich, 8093 Zurich, Switzerland}
\author{Ross~Shillito}
     \affiliation{Institut Quantique and D\'epartement de Physique, Universit\'e de Sherbrooke, Sherbrooke J1K 2R1 QC, Canada}
\author{Paul~Magnard}
    \affiliation{Department of Physics, ETH Zurich, 8093 Zurich, Switzerland}
\author{Ants~Remm}
    \affiliation{Department of Physics, ETH Zurich, 8093 Zurich, Switzerland}
    \affiliation{Quantum Center, ETH Zurich, 8093 Zurich, Switzerland}
\author{Christoph~Hellings}
    \affiliation{Department of Physics, ETH Zurich, 8093 Zurich, Switzerland}
    \affiliation{Quantum Center, ETH Zurich, 8093 Zurich, Switzerland}
\author{Nathan~Lacroix}
    \affiliation{Department of Physics, ETH Zurich, 8093 Zurich, Switzerland}
    \affiliation{Quantum Center, ETH Zurich, 8093 Zurich, Switzerland}
\author{Quentin~Ficheux}
    \affiliation{Department of Physics, ETH Zurich, 8093 Zurich, Switzerland}
\author{Dante~Colao~Zanuz}
    \affiliation{Department of Physics, ETH Zurich, 8093 Zurich, Switzerland}
    \affiliation{Quantum Center, ETH Zurich, 8093 Zurich, Switzerland}
\author{Graham~J.~Norris}
    \affiliation{Department of Physics, ETH Zurich, 8093 Zurich, Switzerland}
    \affiliation{Quantum Center, ETH Zurich, 8093 Zurich, Switzerland}
\author{Alexandre~Blais}
     \affiliation{Institut Quantique and D\'epartement de Physique, Universit\'e de Sherbrooke, Sherbrooke J1K 2R1 QC, Canada}
     \affiliation{Canadian Institute for Advanced Research, Toronto, Ontario M5G 1M1, Canada}
 \author{Sebastian~Krinner}
    \affiliation{Department of Physics, ETH Zurich, 8093 Zurich, Switzerland}
    \affiliation{Quantum Center, ETH Zurich, 8093 Zurich, Switzerland}
\author{Andreas~Wallraff}
    \affiliation{Department of Physics, ETH Zurich, 8093 Zurich, Switzerland}
    \affiliation{Quantum Center, ETH Zurich, 8093 Zurich, Switzerland}
    \affiliation{ETH Zurich - PSI Quantum Computing Hub, Paul Scherrer Institute, 5232 Villigen, Switzerland}

\date{\today}
\begin{abstract}
  The performance of a wide range of quantum computing algorithms and protocols depends critically on the fidelity and speed of the employed qubit readout. Examples include gate sequences benefiting from mid-circuit, real-time, measurement-based feedback, such as qubit initialization, entanglement generation, teleportation, and perhaps most importantly, quantum error correction. A prominent and widely-used readout approach is based on the dispersive interaction of a superconducting qubit strongly coupled to a large-bandwidth readout resonator, frequently combined with a dedicated or shared Purcell filter protecting qubits from decay. By dynamically reducing the qubit-resonator detuning and thus increasing the dispersive shift, we demonstrate a beyond-state-of-the-art two-state-readout error of only \SI{0.25}{\percent} in 100\,ns integration time. Maintaining low readout-drive strength, we nearly quadruple the signal-to-noise ratio of the readout by doubling the readout mode linewidth, which we quantify by considering the hybridization of the readout-resonator and its dedicated Purcell-filter. We find excellent agreement between our experimental data and our theoretical model. The presented results are expected to further boost the performance of new and existing algorithms and protocols critically depending on high-fidelity, fast, mid-circuit measurements.
\end{abstract}
\maketitle

Realizing high-fidelity and fast single-shot readout of a qubit \cite{Mallet2009,Reed2010a,Walter2017} is essential for quantum error correction protocols \cite{Kitaev2003,DiVincenzo2009,Andersen2020b,Krinner2022,Acharya2023} in which qubit decoherence during readout and reset contributes significantly to the logical error. It is also key for algorithms requiring real-time feedback, such as teleportation \cite{Bennett1993, Gottesman1999, Steffen2013, Qiu2023}, distillation \cite{Bennett1996a, Bravyi2005} and initialization \cite{Johnson2012, Riste2012, Salathe2015, Herrmann2022}.

In superconducting circuits, the most commonly used readout architecture employs the state-dependent dispersive shift of the resonance frequency of a resonator coupled to the qubit to infer the qubit state \cite{Blais2004,Wallraff2005,Blais2021}. Whilst the frequency of the resonator is typically fixed, flux-tunable transmons allow to control the qubit-resonator detuning by modifying the transmon frequency \cite{Koch2007}, enable high-fidelity fast entangling gates \cite{Negirneac2021,DiCarlo2010,Strauch2003} and avoid frequency collisions. Additionally, each qubit is often coupled to a microwave transmission line via a dedicated \cite{Heinsoo2018,Andersen2020b,Krinner2022} or common Purcell filter \cite{Arute2019,Chen2021p,Acharya2023} to protect the qubit from radiative decay \cite{Reed2010,Jeffrey2014,Bronn2015b}. Such measurements are usually performed with \textit{weak} measurement tones to avoid nonlinearities and detrimental qubit state transitions, although high-power readout has been studied both theoretically \cite{Boissonneault2010} and experimentally \cite{Reed2010a}.

In the past few years, significant improvements to the single-shot readout have been realized, reaching a two-level readout assignment fidelity of $4\times 10^{-3}$ in 88\,ns~\cite{Walter2017}. Faster readout protocols have been realized, with a $9\times 10^{-3}$ fidelity readout achieved in 40\,ns by utilizing the distributed-element, multimode nature of the readout resonator~\cite{Sunada2022}.

One of the critical parameters governing dispersive qubit readout is the detuning between the qubit and the readout resonator, which controls both the magnitude of the dispersive shift and the nonlinearities induced in the resonator. Different detuning regimes have been explored, including cases where the resonator frequency is \textit{lower} than the qubit \cite{Walter2017,Khezri2022}. Notably, the measurement fidelity has been shown to improve for smaller detunings \cite{Malekakhlagh2022a, Krinner2022}, although these observations were not fully explained.

\begin{figure}[ht]
  \includegraphics{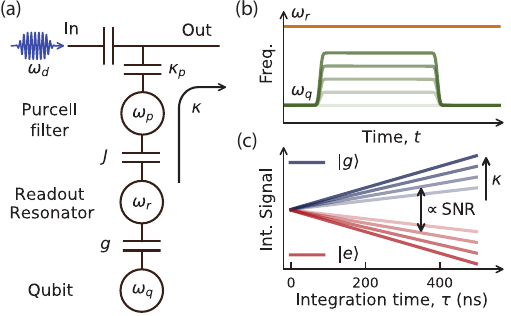}
  \caption{\textbf{(a)} Schematic of a qubit coupled to a readout resonator-Purcell-filter system. The qubit of transition frequency $\omega_q$ is coupled capacitively at rate $g$ to a readout resonator of frequency $\omega_r$. The readout resonator in turn is coupled at rate $J$ to a Purcell filter of frequency $\omega_p$, which is coupled to a feedline at rate $\kappa_p$. We probe the system by measuring the transmission of a readout pulse at frequency $\omega_d$ through the feedline. The effective decay rate of the readout resonator is indicated as $\kappa$. \textbf{(b)} Schematic of time-dependence of qubit frequency $\omega_q$ relative to the readout resonator frequency $\omega_r$.  The qubit, initially idling at the lower flux sweet spot, is pulsed to a smaller detuning from the readout resonator using a fast Gaussian-filtered, rectangular flux pulse. \textbf{(c)} Illustration of the rise of the SNR of the qubit readout with integration time $\tau$ parameterized by the effective linewidth of the readout resonator $\kappa$ at approximately constant dispersive shift $\chi$. Increasing color saturation in panels (b) and (c) indicate increasing $\kappa$ at reduced detuning between $\omega_q$ and $\omega_r$.}
  \label{fig:Figure1}
\end{figure}

In this work, we demonstrate an increase in the signal-to-noise ratio (SNR) and assignment fidelity by bringing the qubit frequency closer to the readout resonator's frequency using a flux pulse, see illustration in Fig.~\ref{fig:Figure1}~(a,b), achieving a minimum two-level readout error of $2.5\times 10^{-3}$ in $100$\,ns. We accredit this remarkable performance not only to an increase in the dispersive shift $\chi$ imparted by the qubit on the cavity, but also to an increase in the effective linewidth of the targeted normal mode response, caused by bringing the Lamb-shifted readout resonator closer to resonance with the Purcell filter, see Fig.~\ref{fig:Figure1}~(c).

\section{\label{sec:Concept}Readout Parameter Characterization}

We perform the experiment with a transmon qubit of transition frequency $\omega_q/2\pi = 4.14$\,GHz at the lower flux sweet spot~\cite{Hutchings2017} and anharmonicity $\alpha/2\pi = -181$\,MHz. It has a lifetime $T_1=$ \SI{30.4}{\us} and is capacitively coupled to a readout resonator with a coupling strength $g/2\pi =224$\,MHz. The readout resonator is coupled to a feedline used for multiplexed readout~\cite{Heinsoo2018} via a dedicated Purcell filter of linewidth $\kappa_p$ and with a coupling strength $J$, see Fig.~\ref{fig:Figure1}~(a). The qubit is located on a device used to execute a distance-three surface code (see Fig.~\ref{fig:device}, Appendix~\ref{ap:device}). Further information on the device properties and its fabrication can be found in Ref.~\cite{Krinner2022}.

To determine the readout parameters as a function of the frequency detuning between the qubit and its readout resonator, we perform pulsed spectroscopy experiments. We first prepare the qubit in the ground state $\ket{g}$ or excited state $\ket{e}$, pulse the qubit to a chosen readout frequency $\omega_q$ using a baseband flux pulse, and probe the readout circuit using a \SI{2.2}{\us} long microwave tone. This duration corresponds to the maximum integration time of our readout electronics (see Appendix~\ref{ap:device}). The flux pulses are Gaussian-filtered rectangular pulses with short rising and falling edges minimizing coupling to two-level systems~\cite{Krinner2022}, see Fig.~\ref{fig:Figure1}~(b).

\begin{figure*}[ht]
  \includegraphics{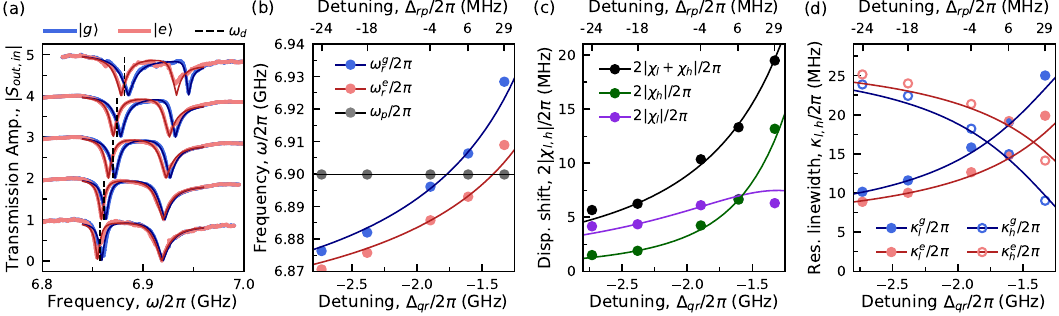}
  \caption{\textbf{(a)} Readout circuitry transmission spectra measured for five qubit---readout-resonator detunings ($\Delta_{qr}/2\pi \in [-2.7, -2.4,  -1.9,  -1.6,-1.3]$\,GHz from bottom to top, vertically shifted by increments of one). Spectra are shown for both the qubit prepared in the ground state $\ket{g}$ (blue) and in the excited state $\ket{e}$ (red). Solid lines are fits based on a coupled qubit---readout-resonator---Purcell-filter model (see Appendix~\ref{ap:device}). Dashed black lines indicate the selected readout frequency, chosen such that the response in transmission $|S_{\rm{out, in}}^{\ket{e}} -S_{\rm{out, in}}^{\ket{g}}|$ is maximum. \textbf{(b)} Resonator frequency $\omega_r^{g}$ ($\omega_r^{e}$) conditioned on the qubit being prepared in the ground (excited) state, and Purcell filter frequency $\omega_p$ as a function of $\Delta_{qr}$. \textbf{(c)} Measured dispersive shift $\chi_{l}$ ($\chi_{h}$) of the lower (higher) frequency hybridized readout mode (see green (purple) points), as a function of $\Delta_{qr}$. The two contributions sum up to the bare readout resonator mode dispersive shift (solid black line) $ \chi=\chi_{l} + \chi_{h} \approx \alpha g^2 /\Delta_{qr}^2$. The solid lines (purple and green) are fits based on a qubit---readout-resonator---Purcell-filter model (see Appendix~\ref{ap:device}). \textbf{(d)} Measured effective readout resonator linewidth $\kappa_{l}$ ($\kappa_{h}$) for the lower (higher) frequency hybridized readout mode for the qubit prepared in the ground/excited (see blue/red), as a function of $\Delta_{qr}$. Solid lines are fits based on a qubit---readout-resonator---Purcell-filter model (see Appendix~\ref{ap:device}). In (b,c,d) the corresponding detunings $\Delta_{rp}=\omega_r^g - \omega_p$ between the Purcell filter and the readout resonator, indicating the degree of hybridization of the two resonator modes, are shown on the top axis.}
  \label{fig:Figure2}
\end{figure*}

We repeat the experiment for five different qubit---readout-resonator detunings $\Delta_{qr}/2\pi$, spanning $-2.7$\,GHz to $-1.3$\,GHz, where $\Delta_{qr} = \omega_q - \omega_r^{g}$. We denote $\omega_r^{g/e}$ as the readout resonator frequency with the qubit prepared in the ground/excited state. The measured (light colored lines) and calculated (dark colored lines) transmission response is shown in Fig.~\ref{fig:Figure2}~(a), with blue/red lines corresponding to the qubit prepared in the ground/excited state. From a fit to a coupled qubit---readout-resonator---Purcell-filter model (see Appendix~\ref{ap:device} and solid blue and red lines in Fig.~\ref{fig:Figure2}~(a)), we extract the relevant readout parameters for each value of $\Delta_{qr}$. The measured (dots) and calculated (lines) dressed readout resonator frequencies $\omega_r^{g}$ and $\omega_r^{e} = \omega_r^{g} + 2\chi$ are shown in Fig.~\ref{fig:Figure2}~(b) (blue and red circles) as a function of the detuning $\Delta_{qr}$, along with the Purcell filter frequency $\omega_p/2\pi=6.900$\,GHz, which remains constant. The variation in the resonator frequencies $\omega_r^{g/e}$ is due to the Lamb shift  $g^{2}/\Delta_{qr}$ caused by the qubit~\cite{Koch2007}. Furthermore, we extract both a large intended Purcell filter linewidth $\kappa_p/2\pi = 34.5$\,MHz and a large intended coupling strength between the readout resonator and the Purcell filter $J/2\pi = 27.5$\,MHz.

We consider a standard circuit-QED approach to model the transmon---resonator---Purcell-filter system, see Appendix~\ref{theory} for details. In the case of a weak drive $\mathcal E$ applied to the filter mode, the readout resonator and the Purcell filter responses can be considered as linear. As such, the dynamics can be effectively mapped to the equations of motion \cite{Gardiner1985}

\begin{equation}
\begin{split}
\begin{bmatrix}
     \dot{\alpha}^{g/e}\\
    \dot{\beta}^{g/e}
\end{bmatrix} =
&-i\begin{bmatrix}
  \omega_r^{g/e} & J\\
    J & \omega_p  - i \kappa_p/2\\
\end{bmatrix}
\begin{bmatrix}
    {\alpha^{g/e}}\\
    {\beta^{g/e}}
\end{bmatrix}\\
&+\begin{bmatrix}
    0\\
    \mathcal{E}e^{-i\omega_d t}\\
\end{bmatrix},
\end{split}
\label{eqn:CoupledSemiclassicalFieldsmaintext}
\end{equation}

\noindent where $\alpha$ and $\beta$ represent the coherent fields of the readout resonator and Purcell filter, respectively. In the regime $J \approx \kappa_p$, we observe two distinct hybridized readout-resonator---Purcell-filter modes, see Fig.~\ref{fig:Figure2}~(a). We denote these as the \textit{low} and \textit{high} readout modes, respectively, the lowest and highest of the two modes in the transmission spectrum. The frequency and linewidth of these modes can be determined, respectively, from the real and the imaginary part of the eigenvalues of the equations of motion in Eq.~(\ref{eqn:CoupledSemiclassicalFieldsmaintext}) in the absence of a drive:

\begin{equation}
\begin{aligned}
        \label{eq:omega_l_h}
        \omega^{g/e}_{{l,h}} &= \frac{\omega_{r}^{g/e}+\omega_p}{2} \pm \frac{1}{2}\Re\sqrt{\left(\Delta^{g/e}_{rp}+\frac{i\kappa_p}{2}\right)^2 + 4J^2},\\
        \kappa^{g/e}_{{l,h}} &=  \frac{\kappa_p}{2} \mp \Im \sqrt{\left(\Delta^{g/e}_{rp}+\frac{i\kappa_p}{2}\right)^2 + 4J^2}.
\end{aligned}
\end{equation}

The readout-resonator---Purcell-filter hybridization leads to a \textit{distribution} of the total qubit-induced dispersive shift $\chi$ on the readout-resonator---Purcell-filter system. Using the model in Eq.~(\ref{eq:omega_l_h}), we can extract the dispersive shifts of the \textit{low} and \textit{high} modes respectively, $\chi_{l/h} = (\omega_{l/h}^{e} - \omega_{l/h}^{g})/2$, see purple (green) circles for the \textit{low} (\textit{high}) mode in Fig.~\ref{fig:Figure2}~(c). While the total dispersive shift $\chi = \chi_l + \chi_h$ shows the expected $\alpha g^2 /\Delta_{qr}^2$ dependence in Fig.~\ref{fig:Figure2}~(c) (solid black line), with $2\chi/2\pi \in [-5.67, -19.49]$\,MHz, the \textit{low} mode dispersive shift only shows small variations in that range, staying between $2\chi_l/2\pi$ $\in [-4.17, -6.69]$\,MHz (see solid purple line in Fig.~\ref{fig:Figure2}~(c)). In contrast, the \textit{high} mode dispersive shift shows a similar scaling with $\Delta_{qr}$ as the total dispersive shift, with $2\chi_h/2\pi \in [-1.5, -13.18]$\,MHz  (solid green line in Fig.~\ref{fig:Figure2}~(c)).

We observe that the dispersive shift of the \textit{low} mode is dominant for qubit---readout-resonators detunings below $-1.6$\,GHz, after which the dispersive shift of the \textit{high} mode becomes larger. The crossing point where $\chi_l = \chi_h$, in the vicinity of the qubit---readout-resonator detuning $\Delta_{qr}/2\pi = -1.6$\,GHz, corresponds to an equal hybridization of the two readout modes. It coincides with $\omega_p = \omega_r^{g} + \chi = \omega_r^{e} - \chi$ being equidistant to the ground and excited state responses of the readout resonator, see Fig.~\ref{fig:Figure2}~(b).

Our model also gives us valuable information about the linewidth of the low and high modes, for the qubit prepared in the ground $\ket{g}$ or excited state $\ket{e}$, namely $\kappa_{l}^{g}$ ($\kappa_{h}^{g}$) and $\kappa_{l}^{e}$ ($\kappa_h^{e}$), as a function of the qubit---readout-resonator detuning $\Delta_{qr}$. As shown in Fig.~\ref{fig:Figure2}~(d), while $|\chi_l| > |\chi_h|$ for $\Delta_{qr}/2\pi \le -1.6$\,GHz, we have $\kappa_{l}^{g/e} < \kappa_{h}^{g/e}$. This is expected as for $\Delta_{qr}/2\pi \le -1.6$\, GHz, the \textit{low} mode has a larger weight in the readout resonator. The difference between $\kappa_{l}^{g}$ and $\kappa_l^{e}$ for the \textit{low} mode derives from the frequency detuning $\Delta_{rp}^{g/e} =\omega_{r}^{g/e}-\omega_p$, between the readout-resonator frequency  and the Purcell-filter frequency. In particular, for the \textit{low} mode, $\kappa_{l}^{g} > \kappa_{l}^{e}$ for all detunings while for the \textit{high} mode $\kappa_{h}^{g} < \kappa_{h}^{e}$, which can be seen from the analysis of the normal mode Hamiltonian in Appendix~\ref{theory}.

In the vicinity of the detuning leading to an equal hybridization of the \textit{low} and \textit{high} modes $\Delta_{qr}/2\pi \approx -1.6$\,GHz, we further note that all $\kappa^{e}_{h}/2\pi \approx \kappa^{g}_{l}/2\pi \approx 19$\,MHz and $\kappa^{g}_{h}/2\pi \approx \kappa^{e}_{l}/2\pi \approx 14$\,MHz. After this crossing point, we observe that while $|\chi_l| > |\chi_h|$ for $\Delta_{qr}/2\pi \ge -1.6$\,GHz, we find $\kappa_{l}^{g/e} > \kappa_{h}^{g/e}$, which we exploit in Sec.~\ref{sec:SNR}. The detailed parameters are summarized in Table~\ref{tab:table1}.
\section{\label{sec:SNR}Single-Shot Readout}
We perform single-shot readout for the qubit---readout-resonator detunings $\Delta_{qr}/2\pi \in [-2.7, -2.4,  -1.9,  -1.6,$ $-1.3]$\,GHz (see Fig.~\ref{fig:Figure2}) as a function of the readout-pulse power and integration time $\tau$ $\in$ [50, 100, 200, 300, 400]\,ns. Each experiment consists of $10^{4}$ single-shot measurements with the qubit prepared in the ground or excited state. The detuning is varied by tuning the qubit to a chosen frequency $\omega_q$, using a flux pulse as described above. We use a rectangular readout pulse with a duration of 450\,ns convolved with a Gaussian filter of width $\sigma=0.5$\,ns, and integrate the readout signal for a time $\tau$ using mode-matched weights \cite{Gambetta2007} to discriminate the ground $\ket{g}$ and excited $\ket{e}$ qubit state responses. The flux pulse lasts longer than the readout pulse. In addition, we use a preselection readout to reduce residual excited state population of the qubit to below 0.1\%~\cite{Magnard2018}.

We express the readout power as a function of the number of photons in the readout resonator $n_g$ when the qubit is prepared in the ground state, relative to the critical number of photons in the resonator $n_{\rm crit} = \Delta_{qr}^2/4g^2$~\cite{Blais2004} at a given qubit---readout-resonator detuning $\Delta_{qr}$. By measuring the qubit-induced ac-Stark shift $\Delta_{\rm ac}=2g^2/\Delta_{qr}$ on the readout resonator at $\Delta_{qr}/2\pi=-2.7\,$GHz we can infer the number of photons $n_g$ in the resonator when the qubit is prepared in the ground state, $n_g=\Delta_{\rm ac}/2(\chi_l+\chi_h)$ (see Appendix~\ref{ap:acstark}). The photon number $n_g$ at other detunings and the photon number $n_e$ when the qubit is prepared in the excited state for all detunings are inferred using semi-classical analysis, see Appendix~\ref{theory}.

The readout drive frequency $\omega_d$ is chosen such that the difference in the response in transmission when the qubit is prepared in the ground or excited state $|S_{\rm{out, in}}^{\ket{e}} -S_{\rm{out, in}}^{\ket{g}}|$, is maximum for the \textit{low} mode, see vertical black dashed lines in Fig.~\ref{fig:Figure2} (a). In the theoretical model (see Appendix.~\ref{app:LinearResponse}), this choice corresponds to selecting the drive frequency which leads to the largest steady-state displacement between the coherent $g$-state and $e$-state Purcell-filter-mode responses. This assumes a fixed weak drive power, such that the response is in the linear regime. We found that this choice consistently leads to a stronger resonator response for the qubit being in the excited state $(n_e > n_g)$. We accredit this to a smaller effective linewidth for the excited state for the lower mode, $\kappa_l^e < \kappa_l^g$, see Fig.~\ref{fig:Figure2} (d). We find this to be an appropriate choice of drive frequency, as the Kerr nonlinearity imparted on the resonator is weaker for the excited state than the ground state (see Appendix~\ref{theory}).

\begin{figure}[ht!]
  \includegraphics{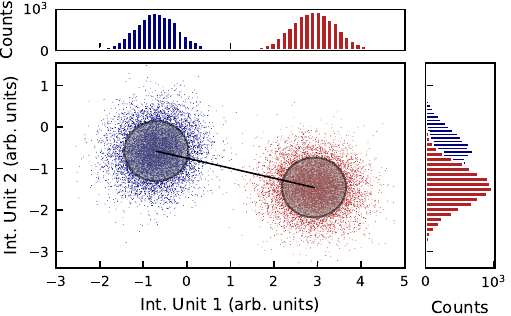}
  \caption{Single-shot readout histogram for a qubit---readout-resonator detuning of $\Delta_{qr}/2\pi=-1.3$\,GHz, a readout integration time of $\tau =$ 100\,ns, and $n_g/n_{\rm crit} =0.93$. We assign the measured state using a bimodal Gaussian mixture model. The marginal distributions of this model along each axis are plotted along the corresponding axis. A solid black line indicates the distance between the means $\mu_g$ and $\mu_e$ of the Gaussian distributions of the ground and excited state responses. The square root of the covariance matrix diagonal elements of the Gaussian distributions $\sigma_g$ and $\sigma_e$ are used as the radii of the black circles.}
  \label{ssro1d}
\end{figure}

We extract the signal-to-noise ratio (SNR) in terms of power, of the acquired single-shot histograms (see Fig.~\ref{ssro1d}) from a bimodal Gaussian distribution as \cite{Gambetta2007}
\begin{equation}
\label{SNR}
\textnormal{SNR} \equiv \left| \frac{\mu_g - \mu_e}{(\sigma_g+\sigma_e)/2} \right|^{2},
\end{equation}

\noindent where $\mu_{g/e}$ and $\sigma_{g/e}$ are, respectively, the mean and the standard deviation of the Gaussian distributions of the $g/e$-state responses. In Fig.~\ref{ssro1d} the solid black line indicates the distance between the means $\mu_g$ and $\mu_e$, and radii of the black circles are given by the square root of the diagonal covariance matrix elements of the bimodal Gaussian distribution.

We characterize the measurement by the average assignment error $\varepsilon_a$ for two-state readout, limited by the overlap error between the Gaussian distributions and the qubit lifetime $T_1$, defined as \cite{Gambetta2007}
\begin{equation}
\begin{split}
\varepsilon_a &= 1 - \mathcal{F}_{g,e} \\
   &= \left[ P(e|g) + P(g|e) \right]/2 \\
   &\gtrsim \frac{1}{2}\left[ 1-\textnormal{erf}\left( \sqrt{\textnormal{SNR} /8} \right) \right] + \frac{\tau}{2T_1},
\end{split}
\label{eqn:T1limit}
\end{equation}
where $P(i|j)$ is the probability of measuring the state $\ket{i}$ when having prepared the state $\ket{j}$, and where the average two-state readout fidelity $\mathcal{F}_{g,e}$ characterizes the quality of the readout. The factor two present in the $T_1$ limit term arises from the fact that only $P(g|e)$ is affected by loss events.

\begin{figure}[ht!]
  \includegraphics{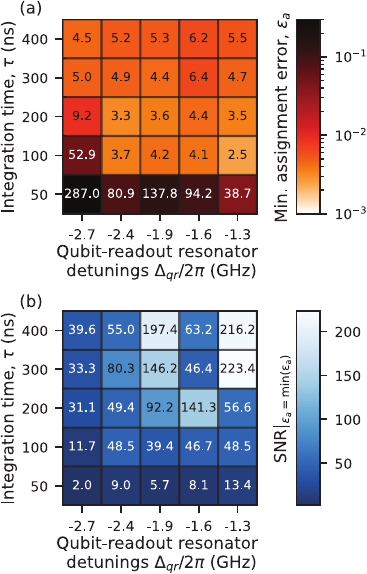}
  \caption{\textbf{(a)} Minimum assignment error $\varepsilon_a$ measured as a function of the qubit---readout-resonator detuning $\Delta_{qr}/2\pi$ $\in [-2.7, -2.4,  -1.9,  -1.6,-1.3]$\,GHz and the readout integration time $\tau \in$ [50, 100, 200, 300, 400]\,ns. Annotated values are in \textit{per mille} unit of probability: the lowest assignment error $\varepsilon_a = 2.5 \times 10^{-3}$ is reached at a qubit---readout-resonator detuning of $\Delta_{qr}/2\pi=-1.3$\,GHz, for an integration time of $\tau=100$\,ns. \textbf{(b)} Measured SNR corresponding to the minimum assignment error in (a).}
  \label{Figure3}
\end{figure}

In Fig.~\ref{Figure3} (a) we present the lowest measured average assignment errors $\varepsilon_a$ as a function of the qubit---readout-resonator detuning $\Delta_{qr}/2\pi \in [-2.7,$ $-2.4,  -1.9,  -1.6,-1.3]$\,GHz and as a function of the readout integration times $\tau$ $\in$ [50, 100, 200, 300, 400]\,ns. We observe that $\varepsilon_a < 1 \times 10^{-2}$ for $\tau \ge 100$\,ns. When $\tau \ge 100$\,ns and for all qubit---readout-resonator detunings, the variations in the average assignment error are small and stay between $2.5 \times 10^{-3} \le \varepsilon_a \le 1\times 10^{-2}$, except for $\Delta_{qr}/2\pi=-2.7$\,GHz and $\tau=100$\,ns. The best assignment error $\varepsilon_{a}=2.5 \times 10^{-3}$ is reached at $\tau=100$\,ns and $\Delta_{qr}/2\pi=-1.3$\,GHz.

This observation suggests that beyond this integration time the assignment error is no longer limited by the SNR, which would continue to increase for longer readout times. This is further demonstrated in Fig.~\ref{Figure3} (b), where we indicate the measured SNR corresponding to each lowest measured average assignment error in Fig.~\ref{Figure3}~(a) as a function of the same qubit---readout-resonator detuning and readout-integration-time range. We observe that a SNR $\ge 30$ leads to $2.5 \times 10^{-3} \le \varepsilon_a \le 1\times 10^{-2}$ for $\tau \ge 100$\,ns and for all detunings $\Delta_{qr}$. On the other hand, SNR $\le 14$ leads to a larger assignment error $3.87 \times 10^{-2} \le \varepsilon_a \le 2.87 \times 10^{-1}$. In particular, we find that the best assignment error $\varepsilon_{a}=2.5 \times 10^{-3}$ is reached for SNR = 48.5. An SNR~$\ge 200$ leads to assignment errors on the same order as an SNR~$\approx 50$ (see for example at $\Delta_{qr}/2\pi=-1.6$\,GHz compared to at $\Delta_{qr}/2\pi=-1.3$\,GHz, with $\tau \in$ [300, 400]\,ns).

\begin{figure}[ht]
  \includegraphics{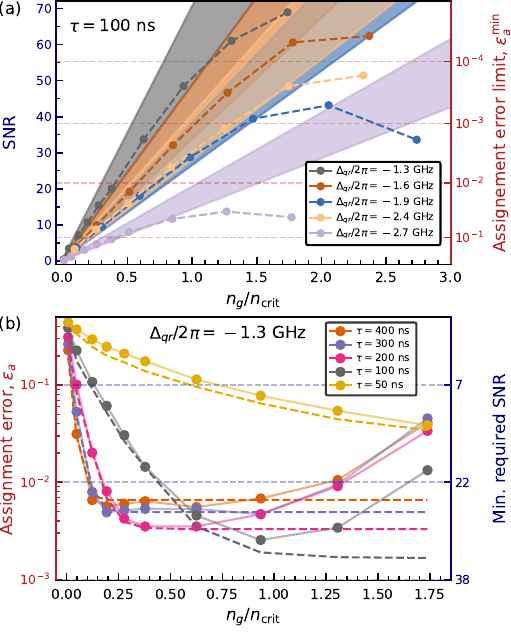}
  \caption{\textbf{(a)} SNR as a function of $n_g/n_{\rm crit}$ for the indicated qubit---readout-resonator detunings $\Delta_{qr}$ and at a fixed readout-integration time of $\tau = 100$\,ns. The shaded regions provide estimates from the analytical solution in the linear regime, detailed in Appendix~\ref{app:LinearResponse}. \textbf{(b)} Average assignment error $\varepsilon_a$ as a function of $n_g/n_{\rm crit}$ for the indicated readout integration times $\tau$ at a fixed qubit---readout-resonator detuning $\Delta_{qr}/2\pi = -1.3$\,GHz. Solid lines are plotted for ease of visualization. Dashed lines correspond to the theoretical limit imposed by the relaxation time of the qubit given by $\varepsilon_a^{\rm min} = 0.5[1-\rm{erfc}(\sqrt{SNR/8})]$ $+\ \tau/2 T_1$.}
  \label{fig:Figure3}
\end{figure}

We next compare the readout performance in terms of SNR at different qubit---readout-resonator detunings as a function of the readout power $n_g/n_{\rm crit}$, for a fixed integration time $\tau=100$\,ns, see Fig.~\ref{fig:Figure3}~(a). The shaded regions indicate the theoretical SNR prediction from the linear response to the readout drive power \cite{Bultink2018}
\begin{equation}
    \textrm{SNR}(t) =2\eta\kappa_p\int_0^t |\beta_e(t') - \beta_g(t')|^2 dt',
\end{equation}
\noindent where $\eta$ is the measurement efficiency. This expression is in line with Eq.~(\ref{SNR}). We note that the SNR for the smallest detunings $\Delta_{qr} \in [-1.6, -1.3]$\,GHz is significantly higher, which we accredit to the increase in the linewidth of the targeted lower mode $\kappa^{g/e}_{l}$, see Fig.~\ref{fig:Figure2}~(d). This increased linewidth results in the pointer states $\beta_{g/e}(t)$ reaching the steady state faster, thus maximizing the SNR rate. The shaded region contains the upper- and lower- bound estimates of the SNR based on uncertainties in the model parameters, see Appendix~\ref{app:LinearResponse}.

In all instances we observe a saturation of the SNR at a readout power $n_g \gtrsim n_{\rm crit}$, where the dispersive approximation is known to break down \cite{Johnson2011,Minev2019,Boissonneault2010}. This is in part due to the broadening of the pointer states caused by the qubit-induced Kerr nonlinearity of the resonator (see Fig.~\ref{fig:Appendix-overlap-error}), measurement-induced state transitions \cite{Sank2016} and ionization \cite{Shillito2022}.
\begin{figure}[h!]
  \includegraphics{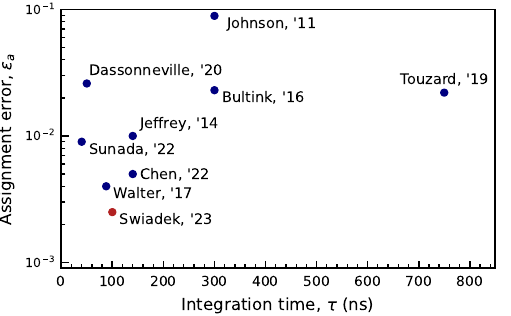}
  \caption{Two-level average assignment error reported in Johnson \textit{et al.} \cite{Johnson2011}, Jeffrey \textit{et al.} \cite{Jeffrey2014}, Bultink \textit{et al.} \cite{Bultink2016}, Walter \textit{et al.} \cite{Walter2017}, Dassonneville \textit{et al.} \cite{Dassonneville2020}, Touzard \textit{et al.} \cite{Touzard2019} (blue circles) and in this work (red circle) as a function of the readout integration time. Jurcevic \textit{et al.} \cite{Jurcevic2021} reached a two-level assignment error of $3.5\times 10^{-2}$ using the excited state promotion technique \cite{Elder2020} and is not plotted here.}
  \label{fig:ref}
\end{figure}

Finally, we compare in Fig.~\ref{fig:Figure3}~(b) the average assignment error $\varepsilon_a$ at different readout integration times $\tau$ $\in$ [50, 100, 200, 300, 400]~ns for a fixed qubit---readout-resonator detuning $\Delta_{qr}/2\pi = -1.3$\,GHz as a function of the readout power $n_g/n_{\rm crit}$. For $n_g/n_{\rm crit} < 1$, we find excellent agreement between the experimental data (dots) and the approximate theoretical limit (dashed lines) in Eq.~(\ref{eqn:T1limit}). Here, we note that the $50\,$ns measurement is clearly limited by the SNR, which consistently improves at higher drive powers. For $100$\,ns, the minimum assignment error $\varepsilon_a^{\rm min} = 2.5\times 10^{-3}$ is limited by the intrinsic lifetime of the qubit $T_1$ rather than the SNR, which can be seen by comparison to the calculated assignment error (gray dashed line), which plateaus at higher readout powers. We notice a distinct upturn in the $100$\,ns measurement at higher drive powers, which we attribute to non-linearities and measurement-induced transitions. 

For the longer readout times $\tau$ $\in$ [200, 300, 400]~ns, the minimum assignment error is obtained at \textit{lower} drive powers, since these measurements reach a larger SNR. Given that the assignment fidelity is limited by the qubit lifetime, the increase in SNR by increasing drive power has little impact on the final assignment error $\varepsilon_a$, as indicated by the plateaus (dashed lines).

\section{\label{sec:Conclusion}Conclusion}

We have demonstrated beyond-state-of-the-art single shot readout reaching a minimum assignment error of $2.5\times 10^{-3}$ in only $100$\,ns when reducing the qubit detuning from the resonator by applying a flux pulse, see our work in perspective with other techniques in Fig.~\ref{fig:ref}. We provided new insights on dispersive readout for a qubit---readout-resonator---Purcell-filter system, in a strongly hybridized regime where the coupling strength between the readout resonator and the Purcell filter $J$ is comparable to the Purcell filter linewidth $\kappa_p$. We showed that by probing the dispersive regime \textit{via} flux pulses we can increase the effective decay rate of the targeted readout mode, thus allowing us to reach larger SNR in a shorter integration time.

Our findings open opportunities to study other regimes and help optimize the readout parameters in the design stage of quantum processors in order to adjust the effective decay rate of the readout mode depending on the applications. For instance, we expect this work to help reduce the readout contribution to the quantum error correction cycle time on superconducting qubit platforms~\cite{Krinner2022}, without compromising on the readout fidelity constraints. Such techniques combined with machine learning methods as in Ref.~\cite{Lienhard2022} for the optimization of pulse shapes, could continue to decrease readout times while maintaining low readout errors.

\section*{\label{sec:ack}Acknowledgements }
The team in Zurich thanks Johannes Herrmann and Stefania La\u{z}ar for contributions to the experimental setup. The team in Sherbrooke thanks Crist\'obal Lled\'o and Catherine Leroux for insightful discussions.

The team in Zurich acknowledges financial support by the Office of the Director of National Intelligence (ODNI), Intelligence Advanced Research Projects Activity (IARPA), through the U.S Army Research Office grant W911NF-16-1-0071, by the EU Flagship on Quantum Technology H2020-FETFLAG-2018-03 project 820363 OpenSuperQ, by the National Center of Competence in Research 'Quantum Science and Technology' (NCCR QSIT), a reseach instrument of the Swiss National Science Foundation (SNSF, grant number 51NF40-185902), by the SNSF R'Equip grant 206021-170731, by the EU-programme H2020-FETOPEN project 828826 Quromorphic and by ETH Zurich. S.K acknowledges financial support from Fondation Jean-Jacques et F\'elicia Lopez-Loreta and the ETH Zurich Foundation. The team in Sherbrooke acknowledges the financial support by NSERC, the Canada First Research Excellence Fund, and the Minist\`ere de l’\'Economie et de l’Innovation du Qu\'ebec. Support is also acknowledged from the U.S. Department of Energy, Office of Science, National Quantum Information Science Research Centers, Quantum Systems Accelerator. The views and conclusions contained herein are those of the authors and should not be interpreted as necessarily representing the official policies or endorsements, either expressed or implied, of the ODNI, IARPA or the U.S Government.\\

The authors declare no competing interests.

\section*{Author contributions}

S.K., P.M. and F.S. planned the experiments with support from all co-authors, and S.K., F.S. performed the experiments. F.S., R.S., and S.K. analyzed the data. R.S. and F.S. worked on the theory. F.S. and A.R. designed the device, and A.R., S.K., D.C.Z. and G.J.N. fabricated the device. C.H., N.L. and A.R. developed the experimental software framework. S.K., A.R., F.S., C.H. and N.L. contributed to the experimental setup and maintained it. F.S., R.S. and S.K. prepared the figures for the manuscript and S.K., A.W., A.B. and Q.F. provided feedback. F.S. and R.S. wrote the manuscript with inputs from all co-authors. A.W., S.K. and A.B. supervised the work.

\appendix
\begin{figure}[t!]
  \includegraphics{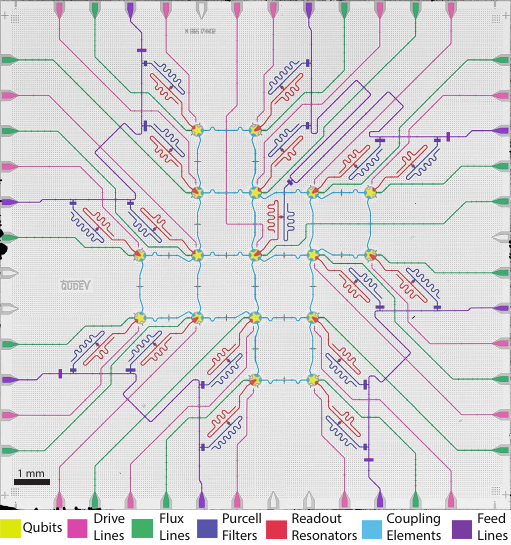}
  \caption{False-colour micrograph of the 17-qubit device used for the experiment, adapted from Ref.~\cite{Krinner2022}; the scale bar denotes 1\,mm. The experiment was realized using the qubit in the feedline at the bottom left corner, as presented in Fig.~\ref{fig:Figure1}.}
  \label{fig:device}
\end{figure}
\section{Experimental setup and device characterization}
\label{ap:device}

We used a qubit of a 17-qubit quantum device, shown in Fig.~\ref{fig:device}, to perform the experiment. We fabricated the 17-qubit quantum processor by sputtering a niobium 150-nm-thin film onto a high-resistivity intrinsic silicon substrate. All coplanar waveguides, capacitors and qubit islands were patterned using photolithography and reactive-ion etching. The aluminium-titanium-aluminium trilayer airbridges establish a well-connected ground plane on the device and connect signal lines split by crossings. We fabricated aluminium-based Josephson junctions using shadow evaporation of aluminium through a resist mask defined by electron-beam lithography.

We characterized the properties of the qubit using spectroscopy and standard time-domain measurements. The qubit has an idling frequency $\omega_q/2\pi =$~\SI{4.144}{GHz}, an anharmonicity $\alpha/2\pi=-181$\,MHz, a lifetime $T_1=$ \SI{30.4}{\us}, a Ramsey decay time $T_2^{*}=$ \SI{29.2}{\us}, and an echo decay time $T_2^{e}=$ \SI{33.9}{\us}.

\begin{table*}[t!]
\begin{ruledtabular}
\begin{tabular}{cccccccc}
qubit---readout-resonator detuning &$\Delta_{qr}/2\pi$& GHz & -2.7                      & -2.4                      & -1.9                      & -1.6                      & -1.3                      \\ \hline
Qubit frequency during readout &$\omega_{q}/2\pi$        & MHz & 4144                   & 4500                  & 5000                   & 5300                   & 5600                   \\
Bare readout resonator frequency &$\omega_{r, b}/2\pi$        & MHz & 6854.63                   & 6858.02                   & 6857.98                   & 6859.74                   & 6864.86                   \\
Dressed readout resonator frequency &$\omega_r^g/2\pi$          & MHz & 6876.27                   & 6881.98                   & 6896.09                   & 6906.33                   & 6928.43                   \\
Purcell filter frequency &$\omega_p/2\pi$                     & MHz & 6899.86                   & 6899.86                   & 6899.86                   & 6899.86                   & 6899.86                   \\
Readout drive frequency &$\omega_d/2\pi$                     & MHz & 6857.4                   &  6861.2               &    6870.0              &     6874.0               &     6881.6               \\
Qubit readout resonator coupling &$g_{b}/2\pi$                    & MHz & 224.32                    & 205.61                    & 211.49                    & 204.2                     & 205.53                    \\
Qubit charge-readout resonator coupling &$g/2\pi$             & MHz &   284.01                        &         271.40                 &       293.71                    &              292.27             &            302.34               \\
Readout resonator-Purcell filter coupling &$J/2\pi$      & MHz & 27.9                      & 27.9                      & 27.9                      & 27.9                      & 27.9                      \\
\textit{Low} mode linewidth, qubit in $\ket{g}$ state&$\kappa_{l}^{g}/2\pi$ & MHz & 10.16                     & 11.61                     & 15.81                     & 19.07                     & 25.00                     \\
\textit{Low} mode linewidth, qubit in $\ket{e}$ state &$\kappa_{l}^{e}/2\pi$ & MHz & \multicolumn{1}{c}{8.88}  & \multicolumn{1}{c}{10.03} & \multicolumn{1}{c}{12.66} & \multicolumn{1}{c}{14.84} & \multicolumn{1}{c}{19.87} \\
\textit{High} mode linewidth, qubit in $\ket{g}$ state&$\kappa_{h}^{g}/2\pi$ & MHz & 23.86                     & 22.41                     & 18.21                     & 14.95                     & 9.02                      \\
\textit{High} mode linewidth, qubit in $\ket{e}$ state&$\kappa_{h}^{e}/2\pi$ & MHz & \multicolumn{1}{c}{25.14} & \multicolumn{1}{c}{23.99} & \multicolumn{1}{c}{21.36} & \multicolumn{1}{c}{19.18} & \multicolumn{1}{c}{14.15} \\
\textit{Low} mode dispersive shift &$2\chi_{l}/2\pi$                            & MHz & -4.17                     & -4.35                     & -6.11                     & -6.69                     & -6.31                     \\
\textit{High} mode dispersive shift &$2\chi_{h}/2\pi$                            & MHz & -1.50                     & -1.90                     & -4.24                     & -6.64                     & -13.18                    \\
Critical readout resonator photon number &$n_{\rm crit}$      &     & 23.14                     & 19.26                     & 10.42                     & 7.55                      & 4.83
\end{tabular}
\end{ruledtabular}
\caption{List of readout parameters extracted for the qubit---readout-resonator detunings $\Delta_{qr}/2\pi$ spanning -2.7\,GHz to -1.3\,GHz using pulsed-spectroscpy measurements.}
\label{tab:table1}
\end{table*}

\begin{figure}[t!]
  \includegraphics{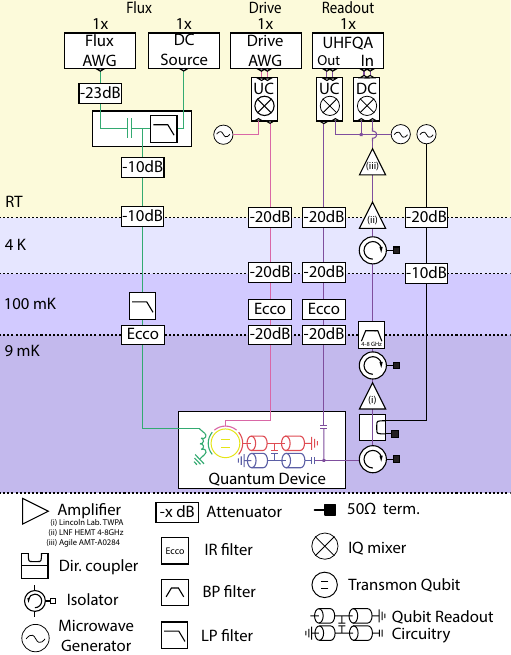}
  \caption{Schematic of the wiring and control electronics. The qubit (yellow) on the quantum device is connected to the room-temperature electronics \textit{via} flux lines (green), drive lines (pink) and readout lines (purple) through its readout resonator (red) and Purcell filter (blue). The background colors indicate to the temperature stages of the experimental setup. }
  \label{fig:fridge}
\end{figure}
Following the method in Ref.~\cite{Heinsoo2018}, we fit the transmission amplitude of the readout signal through a feed-line to the function
\begin{equation}
\begin{aligned}
    &|S_{\rm{out, in}}|(\omega) = (A + k(\omega-\omega_0))\times\\
    &\left|\cos(\phi) - e^{i\phi}\frac{\kappa_p( -2i\Delta_r^{g/e})}{4J^2 + (\kappa_p - 2i\Delta_p)(- 2i\Delta_r^{g/e})}\right|,
\end{aligned}
\end{equation}
where $A$ is the amplitude, $k$ describes a tilt in the spectrum centered at $\omega_0$, $\phi$ is a phase rotation induced by the capacitive couplings to other lines, $\kappa_p$ is the external coupling rate of the Purcell filter,  $\Delta_p = \omega - \omega_p$ is the detuning between the drive frequency $\omega$ and the Purcell-filter frequency $\omega_p$, and $\Delta_r^{g/e}=\omega - \omega_r^{g/e}$ is the detuning between the drive frequency and the resonator frequency conditioned on the state of the qubit. The relevant parameters for the studied qubit at the indicated qubit-resonator detunings are provided in Table~\ref{tab:table1}.

We installed the device in a magnetically-shielded sample holder mounted at the base plate (9\,mK) of a cryogenic measurement setup \cite{Krinner2019} and connected it to the control and measurement electronics as shown in Fig.~\ref{fig:fridge}. We use a DC signal to generate a current inducing a magnetic flux in the SQUID-loop of the transmon qubit, to control its idle frequency.
We use arbitrary waveform generators to apply voltage pulses (2.4\,GSa/s sampling rate) to the qubit to tune its frequency for readout. The DC and AWG signals are combined using a bias-tee. A precompensation of distortions in the flux line is applied, as in Ref.~\cite{Krinner2022}.

We perform the single-shot readout experiments with an ultra-high frequency quantum analyzer (UHFQA) by using an IQ-mixer to upconvert the frequency-multiplexed readout pulses from an intermediate frequency signal sampled at 1.8\,GSa/s to the gigahertz frequency range of the readout circuitry. At the output of the device feedline, the readout signal passes through a wide-bandwidth near-quantum limited traveling-wave parametric amplifier (TWPA) \cite{Macklin2015}, a high-electron-mobility transistor (HEMT) amplifier, and room-temperature amplifiers. It is then down-converted with an IQ-mixer and digitally demodulated and integrated in the UHFQA. Further details on the device fabrication, characterization, and the experimental setup, can be found in~Ref.~\cite{Krinner2022}.

\section{Model}
To model the system, we use the Hamiltonian
\label{theory}
\begin{equation}
    \begin{split}\label{eqn:FullSystemHamiltonian}
        \hat{H}_{trp} & = 4E_c \hat{n}_t^2 -E_J(\Phi)\cos\hat{\varphi}_t \\
        &+ \omega_{r,b} \hat{a}^\dagger \hat{a}- ig(\hat{n}_t-n_g)(\hat{a} - \hat{a}^\dagger)\\
        &+ \omega_p\hat{f}^\dagger\hat{f} +J(\hat{f}^\dagger -\hat{f})(\hat{a}^\dagger - \hat{a}) \\
        & + 2i\mathcal{E}\sin(\omega_d t)(\hat f^\dagger -\hat f),
    \end{split}
\end{equation}
where $\hat{n}_t$ is the charge operator of the transmon, $\hat{a}$ the readout resonator mode creation operator and $\hat{f}$ the Purcell filter mode creation operator. $E_c$ is the charging energy of the transmon, $E_J(\Phi)$ the flux-tunable Josephson energy of the transmon, $\omega_{r,b},\ \omega_p$ the bare resonator and Purcell filer frequencies, and $g, J$ the transmon-resonator and resonator-Purcell coupling rates respectively. $\mathcal E, \omega_d$ are the the drive amplitude and drive frequency. Further, we consider a master equation
\begin{equation}
\label{eqn:mastereqn}
    \dot {\hat \rho} =-i[\hat H_{trp},\hat\rho] +\kappa_p\mathcal{D}[\hat f],
\end{equation}
where $\kappa_p$ is the coupling rate between the Purcell filter and the feedline.
\begin{figure*}[t!]
  \includegraphics{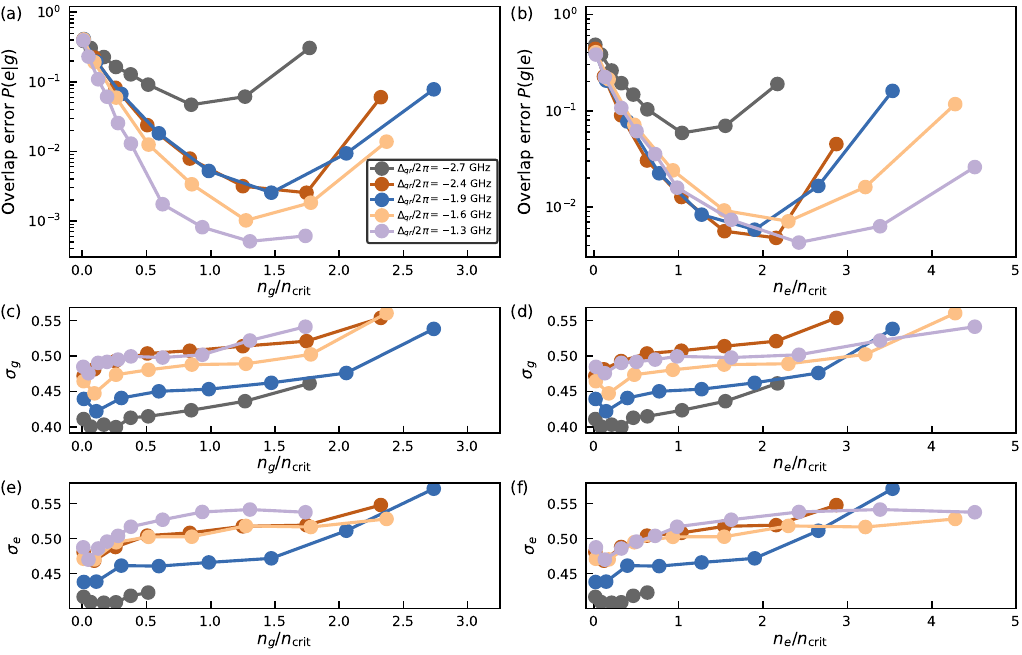}
  \caption{\textbf{(a), (b)} Overlap error $P(e|g)$ and $P(g|e)$ for the indicated qubit---readout-resonator frequency detunings $\Delta_{qr}$ at a fixed integration time of $\tau=100$\,ns. \textbf{(c), (d)} Gaussian width $\sigma_g$ of the ground state single-shot histogram as a function of $n_g/n_{\rm crit}$ and $n_e/n_{\rm crit}$. \textbf{(e), (f)} Gaussian width $\sigma_e$ of the ground state single-shot histogram as a function of $n_g/n_{\rm crit}$ and $n_e/n_{\rm crit}$.}
  \label{fig:Appendix-overlap-error}
\end{figure*}
We first diagonalize the transmon-resonator subsystem. We follow the notation of Ref.~\cite{Blais2021}, and assume a Kerr-nonlinear oscillator model for the transmon, valid in the low readout power regime. A Schrieffer-Wolff transformation yields an effective Hamiltonian
\begin{align}
    \hat{H} &= \bar\omega_q \hat b^\dagger \hat b  + \omega_p \hat f^\dagger \hat f   + \omega_r^g\hat a^\dagger \hat a +2\chi \hat a^\dagger \hat a \hat b^\dagger \hat b \\ \notag
    &-2\lambda'\lambda^3 E_C\hat a^{\dagger 2} \hat a^2 \hat b^\dagger \hat b -\frac{\alpha}{2} \hat b^{\dagger 2} \hat b^2 -\frac{E_c}{2}\lambda^4\hat a^{\dagger 2} \hat a^2\\
    &+ J\left(\left[1 -2\lambda\lambda'\hat b^\dagger \hat b \right]\hat a^\dagger \hat f +\lambda\hat b^\dagger \hat f +\textrm{H.c.}\right),\notag
\end{align}
where
\begin{align}
        \chi &=-g^2E_C/(\Delta_{qr}(\Delta_{qr}-E_C)), \\
        \lambda' &= \lambda E_c/\left[\Delta_{qr} + E_c(1-2\lambda^2)\right]
\end{align}
\noindent with $\lambda = g/\Delta_{qr}$ and $\Delta_{qr}=\omega_q-\omega_{r,b}$. Further, the qubit and resonator frequencies become Lamb-shifted, with  $\bar \omega_q \approx \omega_q + g^2/\Delta_{qr}, \ \omega_r^g \approx \omega_{r,b} - g^2/\Delta_{qr}$. The contribution $ -2\lambda'\lambda^3 E_C\hat a^{\dagger 2} \hat a^2 \hat b^\dagger \hat b$ normalizes down the effective Kerr nonlinearity when the qubit is in the excited state, and we note that $K_e/K_g \approx 1 + 4\lambda'/\lambda$, where $4\lambda'/\lambda < 0$ for $\Delta_{qr} <0$. This nonlinearity leads to a significantly larger increase in the Gaussian width of the ground state response than the excited state response, as seen in Fig.~\ref{fig:Appendix-overlap-error}~(c,d,e,f) \cite{Bartolo2016,Roberts2020}. For this reason, we quote the drive power in the main text as a function of $n_g/n_{\textrm{crit}}$ as opposed to $n_e/n_{\textrm{crit}}$.

We also note a correlation between the broadening of the ground state response and an increase in the overlap errors $P(g|e)$ and $P(e|g)$ for $n_g/n_\textrm{crit} \gtrsim 1$ where nonlinear effects are expected to be more important, see Fig.~\ref{fig:Appendix-overlap-error}~(a,b).
Frequency renormalizations from the effective coupling of the filter to the qubit are on the order of $J^2\lambda^2/\Delta_{qr}^2$ and can be safely ignored in this regime, which was corroborated by numerical diagonalization of Eq.~(\ref{eqn:FullSystemHamiltonian}) and simulation of the master equation.  More importantly, we note that the effective coupling strength between the resonator and Purcell filter, $J[1-2\lambda\lambda'\hat b^\dagger \hat b]$, only weakly depends on the qubit state.
\section{Linear response}\label{app:LinearResponse}
For sufficiently small drive amplitudes, we can assume negligible impact from the Kerr nonlinearity and take the resonator and filter responses to be linear. As such, we can use the relation
\begin{equation}
\begin{split}
    \begin{bmatrix}
    \dot{\alpha}^{g/e}\\
    \dot{\beta}^{g/e}
    \end{bmatrix} = &-i
    \begin{bmatrix}
    \omega_r^{g/e} - \omega_d & J^{g/e}\\
    J^{g/e} & \omega_p - \omega_d - i\frac{\kappa_p}{2}\\
    \end{bmatrix}
        \begin{bmatrix}
    {\alpha^{g/e}}\\
    {\beta^{g/e}}\\
    \end{bmatrix}\\
    &+
    \begin{bmatrix}
    0\\
    \mathcal{E}\\
    \end{bmatrix},
    \end{split}
    \label{eqn:CoupledSemiclassicalFields}
\end{equation}

\noindent where $\alpha$ and $\beta$ represent the coherent fields of the readout resonator and Purcell filter respectively, $\mathcal{E}$ is the drive amplitude and $J^{g/e} = J\left[1 -\lambda\lambda' (\langle \hat\sigma_z \rangle +1)\right]$ is the effective readout-resonator---Purcell-filter coupling, and $\omega_r^e = \omega_r^g + 2\chi$.
\begin{figure}[t!]
  \includegraphics{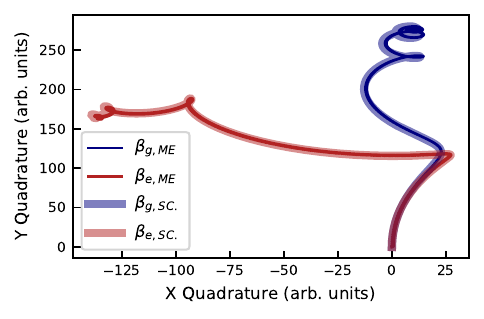}
  \caption{Example comparison of the master equation $\beta_{g/e,ME}$, against the semi-classical trajectories $\beta_{g/e,SC}$, plotted in the phase space of the Purcell filter mode for $\Delta_{qr}/2\pi=-1.3$ GHz and $\mathcal{E}/2\pi = 10$ MHz. Solid lines correspond to the master equation expectations of the resonator and filter modes, $\langle a \rangle$, and $\langle f\rangle$, respectively, where $\langle X \rangle = \Re{a,f}$, $\langle Y \rangle = \Im{a,f}$. Transparent lines correspond to the semiclassical solution. }
  \label{fig:CoherentTrajectories}
\end{figure}
Diagonalizing the equation of motion in the absence of a drive $(\mathcal E = \omega_d=0)$ yields eigenvalues
\begin{equation}
\begin{aligned}
        \lambda^{g/e}_{{l,h}} = &\frac{\omega_{r}^{g/e}+\omega_p -i\kappa_p/2}{2} \\ & \pm \frac{1}{2}\sqrt{\left(\Delta^{g/e}_{rp}+\frac{i\kappa_p}{2}\right)^2 + 4J^{2,g/e}}.
\end{aligned}
\end{equation}
For $4J \gg \kappa$, the eigenvalues approximately correspond to a normal mode, where the indices $l,h$ corresponds to the \textit{low} and \textit{high} mode respectively. In this fashion, the real and imaginary components of $\lambda_{l,h}^{g/e}$ correspond to the frequency and linewidth of these \textit{low} and \textit{high} modes
\begin{equation}
\begin{aligned}
        \omega^{g/e}_{{l,h}} &= \textrm{Re}[\lambda^{g/e}_{{l,h}}], \ \ \kappa^{g/e}_{{l,h}} =  -2\textrm{Im}[\lambda^{g/e}_{{l,h}}].
\end{aligned}
\end{equation}
To obtain a qualitative understanding of the eigenvalues, we perform an expansion of the square root in $(\Delta^{g/e}_{rp}+i\kappa/2)$.
Assuming $J^{g/e}\approx J$, this yields
\begin{equation}
\begin{aligned}
        \lambda^{g/e}_{{l,h}} &= \frac{\omega_{r}^{g/e}+\omega_p -i\kappa_p/2}{2}\\ &\pm \left(J + \frac{\Delta_{rp}^{2ge} -i\Delta_{rp}^{g/e}\kappa_p - \kappa^2/4 }{8J}\right).
\end{aligned}
\end{equation}
Consequently, we see that the frequency $\omega_{l,h}^{g/e}$ of the two sets of normal modes are approximately separated by $2J$, with the relative dispersive shift of each mode $\chi_{l,h}$ between the low and high mode being
\begin{equation}
    \begin{aligned}
    &2\chi_l = \omega^e_l - \omega^g_l \approx \chi -\frac{\Delta_{rp}^g\chi +\chi^2}{2J},\\
    &2\chi_h = \omega^e_h - \omega^g_h \approx \chi +\frac{\Delta_{rp}^g\chi +\chi^2}{2J},\\
    &\kappa_l^{g/e} \approx \frac{\kappa_p}{2} + \frac{\Delta^{g/e}_{rp}\kappa_p}{4J}, \\
    &\kappa_h^{g/e} \approx \frac{\kappa_p}{2} - \frac{\Delta^{g/e}_{rp}\kappa_p}{4J}.\\
     \end{aligned}
\end{equation}
Noting that $\Delta^{e}_{rp} < \Delta^{g}_{rp}$, we see that $\kappa^{g}_h < \kappa^{e}_h$ and $\kappa^{g}_l > \kappa^{e}_l$ for $\Delta^g_{rp} < 0$, and vice versa for $\Delta^g_{rp} > 0$.

The relevant frequencies and linewidths extracted from the normal-mode Hamiltonian in Eq.~(\ref{eqn:CoupledSemiclassicalFields}) are plotted in Fig.~\ref{fig:Figure2}.

The steady state expressions are found to be
\begin{equation}
\label{eqn:SteadyStateCoherent}
  \begin{bmatrix}
    {\alpha_{ss}^{g/e}}\\
    {\beta_{ss}^{g/e}}
    \end{bmatrix} =
\frac{\mathcal{E}}{\Delta^{g/e}(\Delta_p -i\kappa_p/2) - J^2}
    \begin{bmatrix}
    -J\\
    \Delta^{g/e}
    \end{bmatrix}.
\end{equation}
Using this expression, the full time-dependent response takes the form
\begin{equation}
\begin{aligned}
    \beta^{g/e}(t) = \beta^{g/e}_{ss} &-\mathcal{E}\frac{\lambda_h -(\omega_r + \chi\langle \sigma_z\rangle)}{d}\frac{e^{-i(\lambda_h-\omega_d)t}}{(\lambda_h-\omega_d)}\\
    &+\mathcal{E}\frac{\lambda_l -(\omega_r + \chi\langle \sigma_z\rangle)}{d}\frac{e^{-i(\lambda_l -\omega_d)t}}{(\lambda_l-\omega_d)},
\end{aligned}
\label{eqn:FilterResponseEqns}
\end{equation}
\noindent where $d = \sqrt{(-\Delta_{rp} -i\kappa_p/2 - \chi \langle \sigma_z \rangle)^2 + 4J^2}$.

We then use Eq.~(\ref{eqn:FilterResponseEqns}) to express the SNR as~\cite{Bultink2018}
\begin{equation}
    \textrm{SNR}(t) =2\eta\kappa_p\int_0^t |\beta_e(t') - \beta_g(t')|^2 dt',
\end{equation}
\begin{figure}[t!]
  \includegraphics{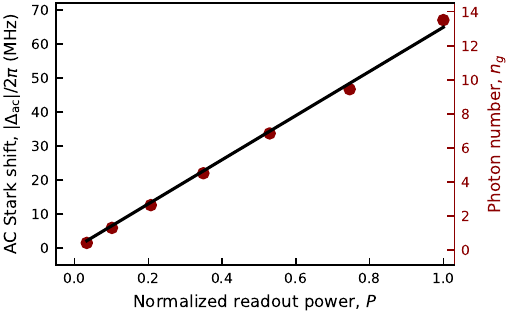}
  \caption{Measured (dots) ac-Stark shift $\Delta_{\rm ac}$ of the qubit prepared in the ground state at $\omega_q/2\pi = 4.14$ GHz as a function of the normalized readout power (real range spans \SI{2.0}{\uV} to \SI{12.5}{\uV}). The corresponding inferred resonator photon number $n_g=\Delta_{\rm ac}/2(\chi_l + \chi_h)$ (solid line) is shown on the right axis.}
  \label{fig:acStark}
\end{figure}
\noindent where $\eta$ is the measurement efficiency. We note that this expression is the square of the often used expression but is in line with Eq.~(\ref{SNR}). We then plot these results for Fig.~\ref{fig:Figure3}~(a) allowing for a $\pm 1$ MHz deviation in the calculated values of $g,\, J,\, \omega_r$ and $\kappa_p$ to allow for uncertainties in the fitted parameters and nonidealities caused by spurious couplings to two-level systems, alongside a variation of up to 5\% in the measurement efficiency at different frequencies. The shaded region contains the upper and low bound estimates of the SNR based on these uncertainties.

Finally, we verify the validity of the semiclassical approximation. Negligible difference was noted in the trajectories in phase space between the expected internal coherent fields $\langle \hat a \rangle, \ \langle \hat f \rangle$, calculated by solving the master equation~(\ref{eqn:mastereqn}), and the corresponding semiclassical predictions $\alpha$ and $\beta$ of the resonator and Purcell filter respectively, confirming that the semiclassical model (in Appendix~\ref{app:LinearResponse}) captures the state separation at low powers. Example trajectories at low power for the Purcell filter mode are plotted for the $\Delta_{qr}/2\pi = -1.3$ GHz case in Fig.~\ref{fig:CoherentTrajectories}.

\section{Photon number and drive power calibration}
\label{ap:acstark}

We measure the ac-Stark shift $\Delta_{\rm ac}$ caused on the qubit prepared in the ground state by the readout resonator as a function of power, see Fig.~\ref{fig:acStark}. To this mean we simultaneously apply a readout tone with a length of \SI{0.6}{\us} of variable power and a $\pi$-pulse of variable frequency. We measure the excited state population as a function of the drive pulse frequency for each readout power for the $\omega_q/2\pi=4.14$\,GHz qubit frequency. The frequency at which the excited state population is maximum corresponds to the instantaneous and ac-Stark shifted qubit frequency.

We determine this frequency using a Gaussian fit. We infer and then calibrate the steady state readout resonator photon number $n_g$ with the qubit prepared in the ground state from $n_g=\Delta_{\rm ac}/2(\chi_l+\chi_h)$ for the chosen drive powers \cite{Schuster2005}. Then, using the steady state resonator response from Eq.~(\ref{eqn:CoupledSemiclassicalFields}), this allows us to extract the effective drive amplitudes $\mathcal{E}$. The steady state resonator responses for the qubit-resonator detunings $\Delta_{qr}/2\pi \in [-4.5, -5.0, -5.3, -5.6]$\,GHz are then inferred from Eq.~(\ref{eqn:SteadyStateCoherent}) at the same drive powers.

\bibliography{bibliography}
\end{document}